Research-based assessment affordances and constraints: Perceptions of physics faculty
Adrian Madsen, Sarah B. McKagan, American Association of Physics Teachers
Matthew "Sandy" Martinuk, Alexander Bell, Cognition Technology
Eleanor C. Sayre, Kansas State University

**Abstract**

To help faculty use research-based materials in a more significant way, we learn about their perceived needs and desires and use this information to suggest ways for the Physics Education Research community to address these needs. When research-based resources are well aligned with the perceived needs of faculty, faculty members will more readily take them up. We used phenomenographic interviews of ordinary physics faculty and department chairs to identify four families of issues that faculty have around research-based assessments (RBA). First, many faculty are interested in using RBAs but have practical needs around how to do so: how to find them, which ones there are, and how to administer them. They want help addressing these needs. Second, at the same time, many faculty think that RBAs are limited and don't measure many of the things they care about, or aren't applicable in their classes. They want assessments to measure skills, perceptions, and specific concepts. Third, many faculty want to turn to communities of other faculty and experts to help them interpret their assessment results and suggest other ways to do assessment. They want to norm their assessment results by comparing to others and interacting with faculty from other schools to learn about how they do assessment. Fourth, many faculty consider their courses in the broader contexts of accountability and their departments. They want help with assessment in these broader contexts. We also discuss how faculty members role in their department and type of institution influence their perceived wants and needs around assessment.

## I. INTRODUCTION

Physics education researchers have produced a plethora of research-based assessment tools[1], but their use by faculty is often limited[2], in part because faculty members don't perceive these materials as meeting their genuine needs. Faculty work within diverse environments, constrained by diverse structures which may not match the environments and structures of materials developers. When developers disseminate materials as finished products, faculty don't feel empowered to use their expertise in teaching and assessing student learning to adapt the materials to the classes in their local contexts[3]. If ordinary faculty are going to use research-based materials in a more significant way, researchers need to learn about their perceived needs, environments, and characteristics. Developers should use this information to create new materials or adapt and frame existing materials. When resources are well aligned with the perceived needs of faculty, faculty members will more readily take them up.

Research-based assessment instruments have had a major impact on physics education reform by providing a universal and convincing measure of student understanding that instructors can use to assess and improve the effectiveness of their teaching. Studies using these instruments consistently show that research-based teaching methods lead to dramatic improvements in students' conceptual understanding of physics[4]. The use of research-based assessment has the potential to transform teaching practice by informing instructors about their teaching efficacy so that they can improve it. Because of this great potential for teaching improvement, in this study, we focus specifically on faculty needs and challenges around assessment.

There are a wide range of ways faculty assess their students' learning and the effectiveness of their teaching. Henderson et al.[2] developed a set of assessment categories through their analysis of 72 interviews with physics faculty about their knowledge and use of research-based instructional methods and associated assessment issues. These assessment strategies are those faculty mentioned using to gauge the effectiveness of their teaching (as opposed to simply understanding what their students learned). Henderson et al.[2] looked at how many of the faculty they interviewed discussed using each of the assessment types. Faculty primarily rely on exams and other assignments (75% of faculty interviewed) as well as informal formative assessment (63%) to gauge the effectiveness of their teaching. They also found about 40% of faculty used student evaluations of teaching or systematic formative assessment. Research-based assessments were used by 33% of the faculty involved in the study. Rarely used assessment strategies include peer observations, teaching portfolios, and post-course feedback. Henderson et al.[2] gives us a sense of how faculty use many kinds of assessment to understand their teaching. In our study, we focus on specifically on research-based assessment, one of the assessment methods discussed in the Henderson study. We go beyond looking at what assessment faculty use and explore faculty's perceived experiences and needs with assessment.

In this empirical study, we investigate the challenges that faculty face around assessment and the types of resources that interest faculty most in order to address these challenges. We also look at how faculty interest varies with the characteristics of the faculty members and their environments. Finally, we make recommendations to the physics education research community on what types of assessments to develop as well as ways to adapt existing resources to more closely align with faculty needs. Our research questions are: 1) What kinds of assessment resources or help with existing assessment interest faculty most? 2) How does this interest vary by type of institution and by faculty member? The goal of this research is to develop categories of the perceived needs of physics faculty around assessment, and to suggest possible solutions for the physics teaching and education research communities to help faculty with their needs.

Using phenomenographic interviews of ordinary physics faculty and department chairs, we identified four families of issues that faculty have around research-based assessments (RBA) and suggest ways that the PER community can address these issues. First, many faculty are interested in using RBAs but have practical needs around how to do so: how to find them, which ones there are, and how to administer them. They want help addressing these needs. Second, at the same time, many faculty think that RBAs are limited and don't measure many of the things they care about, or aren't applicable in their classes. They want assessments to measure skills, perceptions, and specific concepts. Third, many faculty want to turn to communities of other faculty and experts to help them interpret their assessment results and suggest other ways to do assessment. They want to norm their assessment results by comparing to others and interacting with faculty from other schools to learn about how they do assessment. Fourth, many faculty consider their courses in the broader contexts of accountability and their departments. They want help with assessment in these broader contexts.

## II. DESIGN THEORETICAL PERSPECTIVE

This work was done specifically to inform the design of new assessment resources on PhysPort[1], a website that supports physics faculty in implementing research-based teaching and assessment practices in their classrooms. These new assessment resources include an interactive assessment data explorer, guides to specific research-based assessments and general expert recommendations on best practices in assessment. In creating these assessment resources, we used methods from user-interface design to create resources well aligned with genuine needs of our users[5–8]. This process generally begins with interviews of potential users to understand who they are and their goals, needs, and experiences. These findings are then synthesized to guide the entire design process. The finished product meets the real needs or desires of a variety of target users, making it much more likely that the product is actually used.

This perspective on designing with users in mind is also represented in the Push-Pull-Infrastructure model of dissemination[9]. This model comes out of medical communication and describes an approach to creating resources that are "pushed" out to users through systematic efforts, and are also "pulled" in by users who seek out the resources because they meet a genuine need. "Push" dissemination activities should be designed to trigger "pull" by the intended user base. When "pull" is triggered, the uptake of a resource or innovation appears to "just happen". In order to create resources that trigger pull, one must understand the needs and wants of potential adopters, and the influences in the social system of the adopters.

Henderson et al.[3] describes four core strategies for faculty change. These include *disseminating curriculum and pedagogy*, *developing reflective teachers*, *developing policy*, and *developing shared vision*. The most commonly discussed strategies in the literature are *disseminating curriculum and pedagogy* and *developing reflective teachers*. The Push-Pull-Infrastructure model of dissemination bridges these two change strategies to leverage the advantages of each while minimizing the shortcomings. *Disseminating curriculum and pedagogy* is limited because developers don't know who will use their resources, so they have problems meeting real faculty needs. *Developing reflective teachers* relies on encouraging faculty to improve their instruction but is limited by the ability of faculty to do this well within the constraints that they face. The push-pull model overcomes these disadvantages by disseminating resources that faculty want within the constraints they face, so that the dissemination will be more fruitful and faculty will uptake high quality RBAs.

## III. METHODOLOGY

### A. Data collection

In order to learn about faculty members' perceived experiences, needs, and desires around assessment, we used phenomenographic interviews. Phenomenography seeks to uncover the qualitatively different ways a phenomenon is experienced[10–12]; in this case, the qualitatively different ways assessment is experienced by faculty. Phenomenography inherently deals with one's *perception* of their experiences, which is exactly what we want to capture in order to ultimately create new or frame existing assessment resources in a way that meets the *perceived* needs of faculty. Phenomenographic studies usually focus on a small number of participants in order to identify the limited number of ways a phenomenon is experienced. The outcome of this type of analysis is a set of categories that express the researchers' interpretation of the participants' experiences.

We interviewed 10 physics faculty and 14 physics department heads in online interviews (Table 1). Video was used in most. We recruited faculty members from a list of participants from the American Association of Physics Teachers (AAPT) Workshop for New Faculty in Physics and Astronomy held several years earlier. We sent faculty an email invitation to participate. We subsequently received recommendations from current participants and the project staff for other faculty who might be willing to participate and invited them. We recruited department heads from a list of participants from the American Physical Society/AAPT Physics Department Chairs Conference held the previous summer. A sufficient number of these department heads agreed to participate, so no additional invitations were needed. Our interviews with faculty took place in the spring and early summer of 2013 and our interviews with department heads took place in the fall of 2013. For both sets of interviews, our goal in recruiting was to interview people from different types of institutions and backgrounds, and a range of interest in research-based assessment. Our overall sample included eight participants from undergraduate-serving public institutions, seven from undergraduate-serving private institutions and 10 from research-focused institutions. Of these, one was trained in physics education research (PER), 16 were current users of RBAs and 5 were not users of RBAs.

Our sample likely oversamples users of research-based assessments, as Henderson et al.[3] found that only 33% of their sample of the 72 faculty they interviewed used RBAs, while 67% of our participants use these assessments. Our invitation and selection mechanism means that a majority of our participants were people who were willing and enthusiastic about talking to researchers about assessment. Most of our participants were friendly toward PER, though not necessarily trained in PER themselves or using research-based materials. Further, most of them were engaged in substantial thinking about issues around assessment in their courses or departments.

This oversampling of a certain type of faculty member was purposeful, as our goal was to talk to faculty who had experienced and reflected on using RBAs so we could understand their perceived wants and needs. Faculty who have already thought about assessment are also more likely to trigger "pull" resources, so designing for them maximizes our impact on physics teaching. Furthermore, department heads are often responsible for setting expectations around assessment, sitting at the interface between the department and the administration as well as thinking more broadly about the department's courses as a whole. For that reason, we sought out department heads to understand how to help them develop policy or shared vision in their departments.

**Table 1.** Distribution of study participants as department heads and faculty members at large and small schools.

|  | Department Head | Faculty |
|---|---|---|
| **Large School** | 9 | 5 |
| **Small School** | 5 | 5 |

The interview protocol for physics faculty and department heads varied slightly (see Appendix for sample protocol). In our interviews with physics faculty, we asked participants about their background, school and department, current teaching practices and use of assessment, needs around assessment, use of research-based assessments, and how our online resources (including general guides to assessment, guides to specific research-based assessment and/or the assessment data explorer) might meet their needs. Our interview protocol for department heads was similar to that used with faculty, except we focused on the way they used assessment at the department level, and not just at the class level.

Two members of our interview team attended each interview, a physics education researcher and a user design experience researcher. During most of the interviews, one team member engaged with the interviewee while the other listened and took notes. The interviews were recorded.

After each interview, the team members individually wrote down the key points they noticed, primarily attending to the user's motivations and goals, tasks that they commonly completed around assessment, attitudes and beliefs, needs and frustrations. These items were often discussed amongst the team members before the next interview. As is common with semi-structured interviews where follow-up questions spark further discussion, each interview discussed a slightly different set of questions. Additionally, as themes began to emerge in our team member discussions, we updated our interview protocol to probe those themes more carefully in future interviews.

After all interviews were completed, every interview was transcribed using the professional transcription service Rev.com. The transcripts of the interviews became the focus of our analysis.

### B. Data analysis

Following a phenomenographic analysis procedure, we identified categories that described faculty members' experiences of their challenges and wants around assessment. Initially, the first author reviewed each interview to become familiar with the transcript set as a whole. During this review process, she identified broad emergent themes to be investigated in subsequent, more detailed analysis. Next she reviewed each interview transcript, usually in one sitting, with the focus of awareness on one particular aspect of the transcript to learn about the details of how each faculty member discussed that aspect. She then reviewed the transcripts three more times with the focus of awareness on how faculty members discussed challenges, wants, and pressures around assessment, respectively. Each time, attention was also paid to how the characteristics of each faculty member and their institution influenced their experience of each aspect investigated.

During the reading, segments of the transcript containing an idea about assessment were isolated, in line with our KiP perspective.

The next step was to create a set of notes for each focused reading that described information perceived to be important in understanding faculty experience with assessment. The ideas in these notes were then sorted into an initial set of categories by identifying similarities and differences in faculty experiences with assessment. Descriptions of the emergent themes in each category were created. When creating the initial categories, segments of the transcript and the associated notes were used, as the notes sometimes lacked depth. These initial categories were then refined and reorganized by the first author who revisited the data and notes. Next, these initial categories were reviewed by and discussed with the last author, who was familiar with the data. After the discussion, the categories were collapsed into four broad families. We counted the number of faculty who discussed ideas in each category. We also counted the number of faculty who discussed ideas in each family of categories. The last author then verified the validity and robustness of the categories by reviewing a random subset of about 30% of the interview data and associated notes. She independently categorized the ideas into the 4 broad categories. The researchers agreed on their categorization of 89% of the data. Both researchers then discussed and further developed the categories and their descriptions. Finally, we augmented the category descriptions with segments of transcript which illuminate and support each description. These segments represent short examples of sentiments and perception shared among faculty in that category; we chose them as exemplars for their brevity and clarity.

## IV. RESULTS: CATEGORIES

The phenomenographic analysis of the interviews produced four families of categories to describe faculty perceptions of challenges and wants around assessment. Table 2 presents each of the four families, the associated sub-categories and the number of faculty that discussed ideas in each category. These categories are discussed in more detail below with illustrative quotes from the interviews. The order in which the categories are presented corresponds to the ease at which the PER community can address faculty challenges around assessment.

**Table 2.** Descriptions of four major categories (in bold) and associated subcategories and number of faculty or department heads represented by each category.

| Category | Description | Faculty in Category (out of 24) |
|---|---|---|
| **Practical Needs** | **Faculty have practical needs around RBAs: how to find them, which ones there are, and how to give them.** | **19** |
| Challenges with giving RBAs | Faculty perceived challenges with giving RBAs such as student not taking tests seriously or not having time to use these assessments. | 13 |
| Collecting and analyzing results | Faculty want a system or tools to help automate collecting and analyzing RBAI data. | 6 |
| **Limited Measure** | **Faculty think that RBAs are limited and don't measure many of the things they care about, or aren't applicable in their classes. Instead they want assessments to measure skills, perceptions and specific concepts.** | **23** |
| Limitations of understanding | Faculty feel that RBAs can only tell you some of what students understand. | 11 |
| Alignment with Course | Faculty feel that the content assessed by RBAs does not align well with the content in their course. | 5 |
| Difficulty with results interpretation | Faculty have trouble interpreting their RBAI results because of the small numbers of students in their course and large fluctuations from year to year. | 14 |
| Want assessments of skills | Faculty want to assess skills like critical thinking, problem solving and thinking like a physicist. | 7 |
| Want assessment of concepts | Faculty want new conceptual assessment, primarily for upper division courses. | 5 |
| Want assessment of student perceptions | Faculty want to ask students about their experience in the course, e.g., how much time they spend on it, how they feel about physics, what motivates or engage them etc | 7 |
| **Want Help** | **Faculty turn to communities of other faculty and experts to help them interpret their assessment results and suggest other ways to do assessment. They want to norm their assessment results by comparing to others and interacting with faculty from other schools to learn about how they do assessment.** | **12** |
| Comparison | Faculty want to compare their assessment data to comparable data from other schools to get a sense of how they are doing. | 7 |
| Community | Faculty want to interact with other faculty to learn about assessment and better interpret their results. | 6 |
| **Broader Context** | **Faculty consider their courses in a broader context of their departments and accountability and face issues and want help with assessment in this context.** | **17** |
| Program-level assessment | Faculty want to assess students' progress through and at the end of their programs | 10 |
| Assessment pressures | Faculty face assessment pressures primarily as a result of accreditation requirements | 16 |
| Challenges with faculty buy-in | Faculty felt using RBAs limits their academic freedom. | 5 |

## A. Practical needs

Many faculty are interested in using RBAs, but have faced challenges with the practicalities of doing so. They want to know what is available and how to use these assessments. They want help with the practicalities of analyzing results, such as matching students and scoring.

### *1. Challenges with giving RBAs*

Faculty discussed problems with giving concept inventories to their students. There is a lot of variation in the kinds of problems people have when giving RBAs. Many faculty discussed students not taking assessments seriously, perhaps because the students have taken so many assessments or they tests don't feel like a natural part of the course. Other common challenges include not having enough time to find and give tests and not knowing how to give tests or giving incorrectly. This category includes eight department heads and six ordinary faculty; five faculty from small schools and nine from large schools.

> "I guess what I'm most concerned about in many assessments is that they'd not be terribly intrusive and that they'd not feel like an unnatural part of the class. The students can usually figure out when something is not a part of the grade or is extraneous. It's not very closely related to what they've been doing. I find that compliance is not as good then**."** –Department head, small school

> "Yeah. For the FCI, that's the information I got from reading the research papers about it that I did and I appreciated that. But I felt like I didn't know how to use the test until I read all of that. I think having that information available is very important. I like to see those sorts of studies but I almost also appreciate a quick start guide where this is just how you should do it. This is how you should give the exam, this is how much time they should have. That kind of thing. That was hard to find." –Faculty, large school

> "I want to find a better tool for the mechanics courses. Or maybe find a way to mix and match some tools, so I have something specific to our Physics 1 class that covered all the topics, but I cannot really give a 30-question mechanics assessment, a 30-question fluids assessment and a 30-questions thermo assessment for just one course. Both pre and post, it's just not going to work. I would really like to see something that's more easily modularized, things that I can pull." -Department head, large school

### *2. Collecting and analyzing results*

Faculty discussed wanting a system or tools to help automate collecting and analyzing RBAI data. This category includes five department heads, one ordinary faculty; one faculty from a small school and five from large schools.

"I'd like a better way to collect our data and analyze our data. It'd be great if we could have, well, yeah, forty laptops that students could use and do things online, and then I could get the data a little easier and maybe more accurately; that would be nice to have at the school" -Department head, large school

"Yeah a visualization tool, where all I have to do is feed the data and push a button, Yay! Right now as it is, I have to figure out how to program spreadsheets through all this." -Department head, large school

"If this visualization tool was like, oh you're using the FCI, yeah, well you realize you can get full results, but you can get these sub results off these 6 questions are analyzing kinematics, here's your kinematic score. Here's your friction score. All of that." -Department head, large school

### *3. Want information on assessments*

Faculty want information about which assessments are available and information on how to do a certain type of assessment. Faculty commonly requested a central place to find summaries of best practices, access to assessments themselves and guides to creating or using research-based assessment. Faculty particularly want information on assessment for upper division courses. This category includes nine department heads, six ordinary faculty; eight faculty from small schools and seven from large schools.

"Well I would like to know what's available and how to use it, right? We're always looking at ways to get better at what we do and this is just one of them so." -Department head, small school

"This is why I was saying that if there was a website which has … which summarizes or lists all kind of assessment tools which have been proven to work, that would be of great help." -Department head, large school

Interviewer: "Right. What features would you want to see in a national database?" Dept. Head: "Things that work, that is, exercises or approach or whatever it is, and supporting material for it, and then a way for them to use the material in the class, and then compare whatever outcome they're supposed to get in some meaningful way to what others are doing." -Department head, large school

### *4. Solutions from PER Community*

Most faculty challenges and wants around practical needs have the straight-forward solutions: better dissemination of relevant information and availability of simple tools for analysis. PhysPort[1], a website to help faculty with research-based teaching and assessment, is in the process of developing resources to help with this problem. PhysPort contains expert recommendations to assist faculty in using different RBAs, including troubleshooting help. Faculty also want help collecting and analyzing RBA data. PhysPort is developing an assessment data explorer which automatically analyzes and visualizes results from a variety of RBAs. Faculty also want information on what assessments are available and how to use them. PhysPort contains guides to many of the common RBAs with information on administration and analysis. Further, faculty can learn about which RBAs are available and download them. Faculty we interviewed discussed wanting information on which RBAs are available for the upper division. There are a growing number of upper division assessments that faculty are not aware of (e.g. CURRENT[13] and CUE[14]). Better dissemination of upper division assessments in particular would be helpful to faculty.

Not all faculty concerns have easy solutions. While many faculty expressed an interest in assessing a wider range of topics while not taking up too much class time, but these goals cannot be achieved at the same time. Further, while faculty expressed interest in more modular assessments, most RBAs developed by researchers in physics education are only validated for use as whole, and researchers discourage such modular use.

### B. Limited Measure

Faculty don't think that current RBAs measure the things they want to measure. This is because they are limited in what they can assess, aren't always well aligned with courses and the results are difficult to interpret. Faculty want new assessments that better align with what they want to measure, such as skills, student perceptions and specific concepts.

#### *1. Limitations of understanding*

Faculty discussed limitations of what you can learn about student understanding from RBAs. There is a sense that concept inventories can only tell you some of what students understand, e.g., they don't measure problem solving, don't give you a deep understanding of student knowledge etc. Also, faculty feel these tests may not tell you more than you already know with commonly used assessments like homework and exams. This category includes five department heads, six ordinary faculty; six faculty from small schools and five from large schools.

> "No survey that you ask will ever tell you exactly what's going on. As soon as you decide on some questions, you have certain priorities and I may or may not agree that those are the right priorities." -Faculty, large school

> "It seems to me that recently, lots of these tests focus very much on conceptual part and I agree that it's very important if the student doesn't have some sort of feeling for how things are going on then they're not really physicists. I don't think throwing away the numerical part or calculational part is good because they may have a field that the force will go up or down but if they can't calculate the force, I'm sorry, they're not physicists either." -Faculty, large school

> "I remember with the optics diagnostic test, it seemed kind of weird (laughs) ... Not the way that I look at optics. I'm a very optical citizen, especially classic optics, that's my field. One might wonder the origins of some of these exams ... I guess- I certainly like the idea. I think that there might be an over-emphasis on concepts and under-emphasis on problem-solving. I think problem-solving has become a bit of a pejorative that people say, 'Well, they can only solve the problems, but they don't understand what's really going on.'" -Faculty, large school

### *2. Alignment with course*

Faculty discussed problems with research-based assessments being well aligned to their course content. They discussed problems with tests only covering some of the curriculum, containing questions that weren't covered in the course or the tests not covering the material that was covered in the course. This category includes three department heads, two ordinary faculty; two faculty from small schools and three from large schools.

"The idea of concept testing is slippery. It has to be pretty well matched to the curriculum you're delivering. One of the problems I ran into since we are using Matter and Interactions for physics is that these kinds of concept tests really, they don't help. They don't show much of anything." –Department head, small school

"I'm no longer interested in really trying hard to adapt my curriculum so that students can get that particular tricky concept right." -Department head, small school

"We've also done the Force Concept Inventory. However, we have changed a little bit the content; for instance, of Physics 100 to make it more relevant to the students that take this course… We have gone away a little bit from the calculation-heavy parts, let's say like 2-dimension projectile motion, and more towards some non-traditional physics that the students encounter, let's say, in their everyday lives…That's why we don't use the physics Force Concept Inventory as often anymore, because simply it doesn't really apply anymore." –Faculty, large school

### *3. Difficulty with results interpretation*

Faculty discussed problems they've had with interpreting or using the results from RBAs. The biggest problem is that small student enrollments make it difficult to interpret the results because statistical analysis is not valid on these small numbers. In small courses, there are large fluctuations in scores from year to year and it's hard to tell if the fluctuations mean anything. Faculty also discussed wanting a deeper understanding of what the results mean, for example, a better understanding of what a specific score tells them. This category includes ten department heads, four ordinary faculty; nine faculty from small schools and five from large schools.

"It's hard to make much sense out of the data that we get out of those classes because the statistics are terrible. If you go from five students answering the question correctly to seven students answering the questions correctly, can you really say that it improved or is it just a statistical fluctuation? Until you really have a lot of data, I think it's hard to make comments about whether you improve from one semester to the next." –Department head, large school

"I want a clear sense of what happened and what can I do to improve it or what can I do to do it again. It was a good outcome. Either way, I want to know why it worked and how to use the information. Concept tests for me, they don't do that… That's ultimately what I want on the assessments." –Department head, small school

"The most gratifying thing as an instructor is to see how far they've come or how things have progressed, especially if you're looking over multiple years. You can say 'Wow, my students are doing better each year as they go through this.' Because I'm improving in my instruction. I don't know. I look at the statistical numbers and it seems I get too small variation for me to detect. I'm just looking at these forty students. For me to draw very meaningful conclusions from it would be hard." - Department head, small school

*4. Want assessments of skills*

Faculty discussed wanting assessments to measure skills (as opposed to content). Some faculty want to measure skills like problem solving, thinking like a physicist, and critical thinking because they want to show that their courses are helping students develop these skills that are vital to being a physicist or they want to be able to meet learning goals/objectives set by their departments, usually to meet accreditation requirements. Other faculty mentioned want to assess math skills in order to determine if their students were ready for their course. This category includes six department heads, one ordinary faculty; one faculty from small schools and six from large schools.

"I think it will be very exciting to have a better way of measuring just people's ability to think like a physicist, independent of the course material." –Faculty, large school

"What I'm still really looking for is that assessment that says have we taught students how to approach a really hard problem…the kind of problem where you have to struggle with it for a day or two and start reading books and start looking for information and asking people questions and looking around on the web and trying to find an answer and it takes you a couple of days. I don't know how you go about getting that information for how good people are at doing that." –Department head, small school

"I would love to have one assessment that could do critical thinking in any class at any level, non-major freshmen to graduate would be nice." –Department head, large school

### 5. Want assessment of concepts

Faculty discussed wanting specific conceptual assessments. Several faculty want research-based assessments for their upper division courses. A few discussed wanting general conceptual assessments that cover their course content and give them feedback on the effectiveness of their teaching. This category includes four department heads, one faculty; one faculty from a small school and four from large schools.

> "I'd like to see more assessment at the upper division on conceptual understanding but at a deeper level than we assess, say, with the FCI or the CSEM or some of those other conceptual instruments that are typically used in the introductory classes." –Department head, large school

> "Another things is to have … Well we do have this Force Concept Inventory and some other assessment mechanisms but we don't have much for, I think there are … I heard or I listened to some talks where they said there are some things being done for the advanced courses. It would be also nice to have something equivalent for the advanced courses." -Department head, large school

> Dept. Head, large school: "I'd rather see something that maybe has a little more detail in as far as what it's assessing so that you could actually get a better idea on a smaller scale what the student understands."
> Interviewer: "The more detailed assessment, would that be something like the CSEM, FCI or is it another thing?"
> Dept. Head: "Probably it would be something similar to CSEM or an FCI. My sense is to get that kind of detail. You have to actually ask a lot of questions."

### 6. Want assessment of student perceptions

Faculty discussed wanting to ask students about their experience in the course, e.g., how much time they spend on it, how they feel about physics, what motivates or engage them etc. Some feel that there are factors outside of what they do in the classroom that influence their students' learning. This category includes three department heads, four faculty; five faculty from small schools and two from large schools.

> "But one of the things that I wonder, that I see, is that this assumes that a student is very motivated, is actively engaged, and actually doing the work, okay? What I want to know is what is going on outside the classroom? How do they actually do things? How much time do they really spend working? Are they motivated? What's the motivation for them?" –Faculty, small school

> "Yeah, I'm curious about how they feel about physics, especially the premeds. I know that they are pretty scared of physics… Yeah, really I think they feel scared and I think they feel like they are not capable of doing it, even though they definitely are. Knowing what kind of classroom things make them feel intimidated. I would be really interested to know that." –Faculty, small school

"Often times faculty teaching the upper division or the graduate courses, when we get together, we would compare notes about how students are doing and what we find lacking. A common theme that we sort of find among the students is, at least a perception among the faculty is that, the students don't seem to work as hard as we want or expect them to… Part of the thing might be to find out what their expectations are, and whether or not that is in line with what the faculty expects, and perhaps, if they are not, then how do we make our students aware of what they will need to do." –Department head, large school

### *7. Solutions from PER community*

Possible solutions from the PER community around these challenges are complex. Many of the perceived limitations of RBA such as alignment with courses or interpreting results from small courses are genuine problems with using RBAs in certain contexts. To help, the PER community can provide more thorough user education around what these types of assessments measure well and what things they cannot tell you.

Many of the concerns expressed by faculty about the lack of assessments in areas such as problem-solving skills, critical thinking, and upper division physics stem from a lack of awareness of existing assessments on these topics. Most faculty we interviewed were not aware of more than a few RBAs such as the FCI and the CSEM. Some faculty were aware of surveys of attitudes about learning physics, such as the Colorado Learning Attitudes about Science Survey (CLASS)[15] and the Maryland Physics Expectations survey (MPEX)[16]. However, most were *not* aware of existing assessments of problem-solving[17,18], assessments of lab skills[19], rubrics to assess scientific abilities[20], or the growing number of upper division assessments (e.g. CURRENT[13] and CUE[14]). The PhysPort assessments page, which provides information about more than 50 RBAs, will address this lack of awareness.

However, user education and resource dissemination are insufficient to solve faculty's perceived needs in this area, which go beyond what can be tested with existing RBAs. Developers and researchers should work with faculty to create modularized assessments that faculty can adapt to variation in course content, or assessments tuned for use in small classes. Both of these solutions will probably require partnerships between developers and faculty at diverse institutions in order to both meet a perceived need and robustly test the emerging assessments.

More broadly, we need to acknowledge that concept inventories and standardized surveys are a fundamentally limited methodology for measuring student learning. Faculty have noticed. As a research community, we need to think more broadly about other methodologies that are also accessible to ordinary faculty, and find ways to work constructively with instructors to adapt and implement these methods in their classes and departments.

### C. Want help

Faculty want to work constructively with developers and other faculty to learn about and improve assessment in their classes. Faculty turn to communities of other faculty and experts to help them interpret their assessment results and suggest other ways to do assessment. They want to norm their assessment results by comparing to others and interacting with faculty from other schools to learn about how they do assessment. The key difference between this category and the practical needs category is the human connection with outside faculty and experts, and the drive to compare personal data with data from other instructors.

#### *1. Community*

In order to interpret their results, faculty discussed wanting to interact with other faculty around assessment. They primarily want to know what works in other departments and would like some way to access this information, such as a community forum. While most faculty mention wanting to learn from their peers, some would like to actively contribute to a forum, sharing specific feedback about their own and others' materials. Some faculty would like to read case studies about what other departments are doing. This category includes four department heads, three faculty; three faculty from small schools and four from large schools.

> "Yeah it would be kind of neat to see what other people have done under the same ideas, and just, almost a place where I can put mine up there. Others can put theirs up there. Then we could share and chat and someone could go, 'Hey that's stupid. You might want to do this.' Or I could go, 'Yeah mine stinks relative to yours. How did you do that?' It's not so much a handbook but examples. Seeing what other people are doing. A clearinghouse of sharing things like this would be great." – Department head, large school

> "It would be something along the lines like the SPIN-UP report or something like that. So it might have some case studies of departments and what they do so that you can look at what other people do and think about what you're doing and try to adapt that." –Department head, large school

> "Yeah, it would be interesting to me to get contact information somehow or have a public bulletin board where you could post comments or questions about some particular piece of data that was somehow flagged to link up with that particular chunk of data" –Faculty, small school

#### *2. Comparison*

Faculty want a way to compare their students' assessment results to students from other schools in order to interpret their assessment results and improve their teaching. Faculty are interested in benchmarking their scores vs. others'; some want to use this information to adjust the emphases in their instruction, while others are curious about which topics are taught well. This category includes three department heads, three faculty; two faculty from small schools and four from large schools.

> "For example, if our students took that kind of exam [RBA] and we could immediately look at comparisons with other universities and see that we're right on track with all sorts of areas but for some reason our students are doing really well or our students weren't doing very well in some specific thing like special relativity or even just a skill, then we'd know exactly what we need to focus on more in teaching. I think that's the kind of thing that would be really useful." –Faculty, large school

> "We've got all this data. It's a matter of figuring out what to do with it, I guess… The problem with the electromagnetism exam is, like I said, this is where having a better sense of how other schools are doing it, having a database where we can compare at any time how maybe comparable schools are doing which would be very helpful because again it's hard to tell. Or, what does this mean? Can we do it better? Then, in terms of the techniques, it would be interesting to compare different techniques of teaching and how that impacts." –Faculty, small school

> "I think at this point we don't have a lot of information to know whether what we're teaching specifically in our courses is the best. Emphasizing things enough in some cases or if we're emphasizing things too much. I think more data that will allow us to compare what we're doing in our program to other programs would be helpful." – Faculty, large school

### 3. Solutions from PER community

Faculty want to connect with other faculty around RBA, but need help to do so. Faculty want to compare assessment data with one another, preferably linked to teaching methods. A system that could help faculty communicate in general around assessment and enable them to compare data would be most useful. The PhysPort assessment data explorer enables faculty to upload their students' assessment results and compare them to students at similar institutions and the national data set.

Many faculty talked about wanting a human connection, which suggests that dissemination models of change are insufficient to meet these needs. The most effective way to facilitate and sustain these connections both virtually and in-person is an open research question. The PER community can take up this question. Some solutions are available, such as efforts to create and study faculty learning communities[21] and university level science education initiatives where PER trained post-docs work with faculty to improve their teaching with research-based methods and assessments[22,23]. Other, informal efforts include the Global Physics Department[24]. However, more research is needed on the impacts and sustainability of these efforts, particularly for faculty in small departments without many local colleagues.

### D. Broader context

More broadly, faculty want to know how their physics programs are benefitting students beyond learning in a single class. They are curious about learning across the four-year curriculum, and how their programs prepare students for professional life after graduation, or for follow-on courses in other departments. Faculty want to know what works in other departments in order to improve what they are doing. University and accreditation requirements drive assessment of programs in many departments, with departments creating learning goals/objectives and corresponding assessments. However, faculty are also worried that a drive to centralized accountability metrics limits their academic freedom in the classroom.

#### *1. Program-level assessment*

Faculty discussed wanting to assess students' progress throughout and at the end of their programs. Faculty want to know if their programs have helped students learn concepts and gain skills. Several faculty want new metrics to assess what their students have gained at the end of their programs. Several others want to compare their programs to other programs to learn about how to improve. This category includes eight department heads, two faculty; five faculty from small schools and five from large schools.

> "It would be most interesting to see how [students'] abilities have changed over the course of our program." –Department head, small school

> "Mostly I'd just like to know what other departments are doing. That would be really useful because good ideas are hard to come by and it would be good to have them from other people." –Department head, small school

> "I would like to develop some long-term snapshots of students, somewhere. You can imagine giving the FCI pre- and post- at the beginning of the first semester, maybe give it some time during the junior year when they are in the middle of a mechanics class and then maybe give some other test about classical mechanics as they leave…I would like our department to develop a set of metrics that we would sort of just repeatedly use to monitor our students." –Faculty, small school

#### *2. Assessment pressures*

Faculty discussed assessment pressures which primarily stem from accreditation requirements. Some of the faculty felt strong pressure from accreditation bodies and their universities to assess student conceptual learning and skills. These departments often have a set of formal learning objectives or goals that they need to align assessments with. Often, the results of the assessments were not as important as just doing the assessments and reflecting on the results. There are a few departments where the accreditation process is low pressure, but in the case of departments accredited by ABET, the requirements are much more rigorous. This category includes thirteen department heads, three faculty, seven faculty from small schools and nine from large schools.

> "Our university really pushes assessment, and I think it's one of the strengths of our university. We have to do assessment reports. Each program has to have a pretty detailed assessment plan at the university…" –Department head, large school

> "Yeah. They [the accreditation agency] want to see you doing an assessment but it's more than just doing it. They want to see that you're using the results of assessment to evaluate your curriculum. They want to see the whole assessment cycle in action. It's not good enough just collecting data and then throw it on a closet and then never looking at it again. They want to see, okay, our students are not understanding these literacy aspects of quantitative reasoning. We need to evaluate where they're happening in our curriculum and start [inaudible] them out and see what do we need to improve them or how can we improve them." –Department head, small school

> "The external pressures are coming from the … accrediting body for the university. Their main push right now is assessment of gen[eral] ed[ucation] courses and of the gen[eral] ed[ucation] model. They [the external body] haven't been pushing assessment of programs too much. But internally, we are told we must assess our programs." –Department head, large school

### *3. Challenges with faculty buy-in*

Faculty feel that they don't want to be told to use standardized assessments (RBAs). Some feel these assessment limits their academic freedom. Adjuncts or instructors may also resist using RBAs, but may have to because of their position in the hierarchy of the department. This category includes five department heads, zero faculty; two faculty from small schools and three from large schools. Even though no ordinary faculty in our study report in this category, we think this lack is due to our sampling methods, which deliberately looked for faculty amenable to RBA. However, department heads report their faculty may be resistant to impositions from above.

> "…even how one assesses starts to butt up against this idea of academic freedom in some way shape or form. There's a little bit of resistance there perhaps. I think different faculty feel skilled in different ways in terms of their ability to assess specific items." –Department head, large school

> "We have tenured faculty who aren't going to be told that you must do this." –Department head, small school

> "This is something that's, as I said, the intro classes are all part-time instructors. Now I know that if I tell them to do it, they probably will, because they feel that, oh, this is something that the chair is asking, and they may feel reluctant to refuse to do it." -Department head, large school

## *4. Solutions from PER community*

Faculty challenges and wants around the broader contexts of physics departments presents a new set of research agendas for the PER community that involve looking not at the individual course level, but the whole department and program. Faculty want help with understanding how their entire program has benefitted students and also with assessing their programs for accreditation. The PER community can work with faculty and with the assessment and psychometrics community outside of physics to interpret solutions from other fields into physics, and help tune them to the needs of physics departments.

### E. Characteristics of faculty and schools

During the interviews, we noticed that they way faculty discussed their perceptions of assessment was related to their role in the department (faculty or department head) and the size of their institution (small or large). Table 3 shows the distribution of department heads and faculty at large and small school across our four broad categories. The majority of those who discussed issues around the broader context of the their departments were department heads. These issues usually centered around accreditation or university level pressures around assessment, so it follows that department heads would primarily attend to these issues. The distribution of faculty from small and large school who discussed ideas in each category is about even across these broad families of categories; however, some of the sub-categories skew more strongly large or small. Attending to the particular needs of small programs is deeply important to improving the state of physics education because they represent the largest number of physics teaching faculty and about half of all physics majors[25]. However, PER groups tend to cluster at larger schools, and students at large research universities are dramatically oversampled in the PER literature[26].

**Table 3.** Distribution of faculty members and department heads across small and large schools. (out of 10 faculty members and 14 department heads, 14 large schools and 10 small schools.)

|  | **Department Head** | **Faculty Member** | **Large School** | **Small School** |
|---|---|---|---|---|
| **Broader Context** | 14 | 3 | 10 | 7 |
| **Limited Measure** | 13 | 10 | 13 | 10 |
| **Practical Needs** | 10 | 9 | 11 | 8 |
| **Want Help** | 4 | 8 | 6 | 6 |

### V. CONCLUSIONS

This study provides direction for the how the Physics Education Research community can support faculty in using research-based assessment by providing resources well aligned with their perceived needs. We identified four families of faculty perceived needs around RBA, each of which had several sub-categories. All faculty in our study are represented in at least one family of categories. While addressing faculty needs in each category separately is important, a robust solution will necessarily help with multiple perceived needs in multiple categories.

Many faculty use RBAs and want to use them more, and simply need guidance on how to do so effectively. Dissemination efforts such as PhysPort aim to address this need.

On the other hand, many faculty think that RBAs are quite limited in what they can measure. Many feel that their context is sufficiently different and special that RBAs – particularly concept inventories like the FCI – are not a good fit. It is true that RBAs are only one piece of the assessment puzzle, but these instruments are extremely valuable for measuring some aspects of the effectiveness of instruction. The PER community can provide more thorough user education around what these assessments can measure as well as analysis tools to make analyzing and comparing results easier.

The PER community should also develop other useful methods for assessment that meet faculty members' perceived needs. Faculty want to measure aspects of student thinking and experience beyond concepts, e.g. skills like critical thinking or student perceptions of their course experiences. In order to do this, the PER community can provide guidance on existing RBAs and develop new RBAs to meet these needs. Some projects[19,20] are already underway to address these needs, but more are needed.

Faculty are hungry to talk to other faculty about how they do assessment for their courses, programs and accreditation requirements, and to compare their results to those of others within physics but outside their departments. They are don't have time to read research papers about these issues; they want to talk to real people about their particular contexts. PhysPort's Assessment Data Explorer will allow faculty to upload their teaching methods and students' scores for comparison to their peers', addressing their interesting in comparing their results to national standards. However, a better system is needed to facilitate online and face-to-face discussions about these results.

The PER community can help by providing open communication channels so that faculty can discuss their particular assessment issues and needs with a knowledgeable person. Possible mechanisms for facilitating these discussions are diverse and not particularly comprehensive; for example, PER people traveling for other purposes might seek out departments to give colloquia, or PER people might deliberately choose to give talks at conferences frequented by ordinary physicists, or PER people can join public conversations (e.g. Twitter) around issues of teaching and learning. These solutions are all "push" solutions to make information – and people – more available to potential users, thus generating (we hope) more "pull".

Faculty learning communities and on-campus initiatives where physics education researchers work with faculty on their courses are other ways to provide faculty with human interaction around assessment; we can also help facilitate faculty talking to each other around issues of RBA and teaching in general. These efforts are often based at specific institutions and cross disciplinary boundaries to do so. Online faculty learning communities within physics offer the potential to create extended support networks for isolated faculty members who may have local colleagues in physics, or with similar interests in using research-based teaching and assessments.

Almost all faculty in our study are aware of a couple of RBAs for introductory courses, primarily the Force Concept Inventory (FCI)[27] or Conceptual Survey of Electricity and Magnetism (CSEM)[28]. They want information about what other assessments that are available as well as information on how to use them. In addition to providing more human connection around assessment to answer these questions, the PER community can make the information faculty want readily available and advertise its availability. Making this information easily and centrally available across multiple RBAs will increase "pull" by meeting faculty's perceived needs for a clearinghouse around assessment.

Faculty also have perceived concerns and needs around assessment in the broader contexts of their departments. They are facing pressures from their universities and accreditation agencies to assess learning goals/objectives and provide evidence of continuous improvement. It is not clear to many faculty members what they should be assessing and how to do this over the course of their physics program. The PER community can help faculty figure out how to assess those qualities they care about over time and compare to other departments. Physics education researchers can serve as a bridge between ordinary physics faculty and researchers in assessment and student learning more broadly. Partnerships between physics education researchers, physics instructors at diverse institutions, and researchers in assessment and the learning sciences represent exciting new avenues for materials design and development for physics education and basic research on student learning and faculty development.

# References

[1] www.physport.org/assessment

[2] C. Henderson, C. Turpen, M. Dancy, and T. Chapman, " Assessment of teaching effectiveness: Lack of alignment between instructors, institutions, and research recommendations," Phys. Rev. Spec. Top. - Phys. Educ. Res. **10**, 010106 (2014).

[3] C. Henderson, N. Finkelstein, and A. Beach, "Beyond dissemination in college science teaching: An introduction to four core change strategies," J. Coll. Sci. Teach. **39**, 8 (2010).

[4] R.R. Hake, "Interactive-engagement versus traditional methods: A six-thousand-student survey of mechanics test data for introductory physics courses," Am. J. Phys. **66**, 64 (1998).

[5] J. Pruitt and J. Grudin, "Personas: Practice and Theory", in *DUX Proc. 2003 Conf. Des. User Exp.* (ACM, New York, 2003), pp. 1–15.

[6] T. Tullis and W. Albert, *Measuring the User Experience: Collecting, Analyzing, and Presenting Usability Metrics.* (Morgan Kaufmann, 2010).

[7] A. Cooper, *Inmates Are Running the Asylum: Why High-Tech Products Drive Us Crazy and How to Restore Sanity* (Sams - Pearson Education, Indianapolis, 2004).

[8] A. Madsen, M. Martinuk, A. Bell, S.B. McKagan, and E. Sayre, "Personas as a Powerful Methodology to Design Targeted Professional Development Resources ," in *Int. Conf. Learn. Sci.* (Boulder, CO, 2014).

[9] J.W. Dearing and M.W. Kreuter, "Designing for diffusion: how can we increase uptake of cancer communication innovations?," Patient Educ. Couns. **81 Suppl**, S100 (2010).

[10] F. Marton, "Phenomenography," in *Int. Encycl. Educ.*, edited by T. Husen and T.N. Postlewhaite, 2nd ed. (Pergamon, Oxford, UK, 1994), pp. 4424–4429.

[11] F. Marton, "Phenomenography - A research approach investigating different understandings of reality," J. Thought **21**, (1986).

[12] M. Proser, "Some Experiences of Using Phenomenographic Research Methodology in the Context of Research in Teaching and Learning," in *Underst. Phenomenographic Res.* (1994), pp. 31–43.

[13] Baily, M. Dubson, and S.J. Pollock, "Research-based course materials and assessments for upper-division electrodynamics (E&M II)," AIP Conf. Proc. **1513**, 54 (2013).

[14] S. V. Chasteen, R.E. Pepper, M.D. Caballero, S.J. Pollock, and K.K. Perkins, "Colorado Upper-Division Electrostatics diagnostic: A conceptual assessment for the junior level," Phys. Rev. Spec. Top. - Phys. Educ. Res. **8**, 020108 (2012).

[15] W. Adams, K. Perkins, N. Podolefsky, M. Dubson, N. Finkelstein, and C. Wieman, "New instrument for measuring student beliefs about physics and learning physics: The Colorado Learning Attitudes about Science Survey," Phys. Rev. Spec. Top. - Phys. Educ. Res. **2**, (2006).

[16] E.F. Redish, "Student expectations in introductory physics," Am. J. Phys. **66**, 212 (1998).

[17] W.K. Adams, "Chapter 4: Creating a New Problem-Solving Assessment Tool", Dissertation, University of Colorado, Boulder, 2007.

[18] K. Cummings, J.D. Marx, "Beta-Test Data On An Assessment Of Textbook Problem Solving Ability: An Argument For Right/Wrong Grading?," AIP Conf. Proc. 113 (2010).

[19] B.M. Zwickl, N. Finkelstein, and H.J. Lewandowski, "The process of transforming an advanced lab course : Goals, curriculum, and assessments," Am. J. of Phys. **81,** 1 (2013).

[20] E. Etkina, A. Van Heuvelen, S. White-Brahmia, D.T. Brookes, M. Gentile, S. Murthy, D. Rosengrant, and A. Warren, "Scientific abilities and their assessment," Phys. Rev. Spec. Top. - Phys. Educ. Res. **2**, 020103 (2006).

[21] http://www.nsf.gov/awardsearch/showAward?AWD_ID=1431638&HistoricalAwards=false

[22] http://www.cwsei.ubc.ca

[23] http://www.colorado.edu/sei/

[24] http://globalphysicsdept.org

[25] R.C. Hilborn, R.H. Howes, and K.S. Krane, *Strategic Programs for Innovations in Undergraduate Physics: Project Report* (College Park, MD, 2003).

[26] http://meetings.aps.org/Meeting/APR11/Event/146006

[27] D. Hestenes, M.M. Wells, and G. Swackhamer, "Force Concept Inventory," Phys. Teach. **30**, 141 (1992).

[28] D.P. Maloney, T.L. O'Kuma, C.J. Hieggelke, and A. Van Heuvelen, "Surveying students' conceptual knowledge of electricity and magnetism," Am. J. Phys. **69**, S12 (2001).
23

# Appendix

Below is a sample interview protocol for a department head interview. As is common with semi-structured interviews where follow-up questions spark further discussion, each interview discussed a slightly different set of questions.

> I am working on a project to develop resources for physics faculty and department chairs around assessment. The purpose of this interview is to help us learn more about how these resources could address faculty needs around assessment. We will use what we learn from you to improve our resources, and in research studies about faculty use of assessment. On this call today are Sandy and Alex who are also members of the design team. They're just going to be listening in to help inform the design.
>
> If it's ok with you, I'd like to record this session. This will allow me to focus on you and I can then refer back to the recording later if I need clarification. The recording will only be shared with my team members. If we use anything from this interview in publications or presentations, we will do so in a way that does not identify you or your institution.
>
> **Background**
> 1. Tell me about your background and your current position.
>     a. Research area?
> 2. Tell me briefly about your school / department.
> 3. How big are classes?
> 4. Service department or major? How many majors?
> 5. Traditional classes or any student centered reforms?
> 6. What is the mixture of faculty look like?
> 7. Tell me about your role in the department.
>     a. How long have you been in that role?
>
> **Assessment**
> 1. Really broadly, what aspects of student learning in your department do you really care about? (content, attitudes, problem solving) (as department head? as an instructor?)
> 2. How does your department evaluate these?
> 3. What do you do with these results?
>     a. teaching reviews?
>     b. educational improvement / reform?
>     c. other?
> 4. Are you happy with what you're getting from your current methods?
> 5. Have you ever been shocked or delighted by the results of an assessment? Tell me about that time. What happened?
> 6. Have you introduced any new assessments recently?
>     a. Why did you decide to make the change?
>     b. Talk about the details of how you found the new assessment, how you used it, what you did with the results etc.
>     c. How did you know it was successful?
> 7. Are there any specific aspects of your students' thinking or attitudes that your department assesses? (Anything that you do that's special)
>     a. What are they? Why do you assess these?
> 8. Are there aspects of your students' thinking or attitudes that you would like to assess,

but currently do not?
    a. What are they? Why would you like to assess these? Upper division?
9. Are your assessment results used by other elements of the university (i.e. school or university level?)
10. Are you facing any pressures around assessment? External or internal?
11. Is your department involved in any major reform efforts? or assessment efforts?

Current Use of Assessment Data
1. Do you ever compare your instructors' assessment results to other results?
    a. Which results do you compare to?
    b. How do you get access?
2. What kinds of comparison or analysis do you use?
3. What are your goals for these comparisons? What kinds of questions are you trying to answer?
4. What specific tools do you use to make these comparisons?
5. Does your department have learning goals? What are they? How do you assess these?
6. Are there comparisons you would like to make that you don't know how to make?

**Research-Based Assessment**
1. I'm going to ask you about some terminology, I want to ask you about what the implications of these terms so that we understand what they mean when we use them. People often use the term "research-based assessment". What does that term mean to you?
2. What about "concept inventories" or a "validated assessment"? Can provide vague definition here.
3. I'm curious to know what you know about research-based assessment. Can you tell me a little bit of what you know about research-based assessment in physics?
4. What value do you think there (is / would be) in using research-based assessment for your department?
5. Have you ever or do you currently use any kind of research-based assessment?
6. IF YES
    a. Which one(s)?
    b. How did you find them?
    c. How did you choose which one to use?
    d. Can you describe how you use them?
    e. How do you analyze the results?
    f. Interpret the meaning of results?
7. Have you ever experienced any difficulties in accessing research-based assessments?
8. Have you ever experienced any difficulties in administering research-based assessments?
9. Have you ever experienced any difficulties in interpreting the results of research-based assessments?
10. Has there been any other time you've introduced a new assessment or changed how you were doing your existing assessment?
11. IF NO
    a. Have you ever considered using any? What stopped you?

**Guides to Specific Assessments**
One feature of this guide will be a set of guides to specific research-based assessment instruments, including concept inventories and other kinds of assessments such as rubrics for assessing lab skills, teaching observation protocols and open-ended surveys. These guides will include access to instruments and information about the research behind them and how to administer them.

1. If you were considering adopting a new assessment in your department, what information would you need to know in order to judge if that assessment was suitable?
2. How would you judge the quality of a particular assessment?
3. Would you use these guides?
4. What kind of information would you want to see in them?
5. If you wanted to look at or use an instrument, but saw that you had to register or wait for someone to approve a request to access it, how likely would you be to do so?

**General Assessment Guides**
We will be developing guides on best practices for general assessment techniques as part of the PER User's Guide.
1. Are there specific things you would like to know about best practices for writing exams and homework problems?
2. Are there specific things you would like to know about best practices for in class formative assessment?
3. Are there specific things you would like to know about best practices for departmental or program assessment?

**Assessment Results Database**
1. We will be building a database of assessment results so that instructors can compare to others' results. Make comparisons to other schools, conduct basic statistical analyses and comparisons, and visualize your results.
2. Would this database be useful to you? How so?
3. To be clear, using our system would mean changing the way you do things right now, which would certainly be an extra effort. What kind of value could we give you that would make it worthwhile
4. In other words, what can we offer you that you're not getting now? (or make easier/ more efficient)
5. What features would you want to see in it?
6. Would any of the following be valuable to you? (Tell me about the value)
    a. comparing the results of your assessments to other teachers' or departments'?
    b. comparing before and after a change that you make
    c. system matches pre/post data and provides histogram
    d. find another teaching method used with students like yours that is doing really well on a given assessment
    e. using data to make decisions on effectiveness of teaching
    f. make a case for more resources to make changes in teaching methods
7. (for each above) Tell me about the situation where this would be valuable (When in the term? What time of day?)

8. What information about another department would you want to see in order to judge whether its assessment results are comparable to your own?
9. Would you have any concerns about uploading your data? (IRB issues or security of data)
    a. Can you think of anything we could do to address those concerns?

**General**

"Okay we're going to wrap up here are a few final questions for you to think about. "
1. Are there any other features that would help you to access and use research-based assessment?
2. Is there anything else we should have asked about but didn't?